\documentclass[12pt]{article}

\usepackage{graphicx}
\usepackage{dcolumn}
\usepackage{bm}

\newcommand{\nn}{\nonumber}

\newcommand{\be}{\begin{eqnarray}}
\newcommand{\ee}{\end{eqnarray}}
\catcode`\@=11
\def\lsim{\mathrel{\mathpalette\@versim<}}
\def\gsim{\mathrel{\mathpalette\@versim>}}
\def\@versim#1#2{\vcenter{\offinterlineskip
\ialign{$\m@th#1\hfil##\hfil$\crcr#2\crcr\sim\crcr } }}
\catcode`\@=12

\begin{document}

\vspace{-1cm}
\noindent
\begin{flushright}
MPP-2003-77\\
KANAZAWA-03-26

\end{flushright}
\vspace{10mm}
\begin{center}
{\Large \bf
Majorana Phase in Minimal $S_3$ 
Invariant Extension \\ of the Standard Model
}
\vspace*{30mm}\\
\renewcommand{\thefootnote}{\alph{footnote}}
Jisuke Kubo
\footnote{Permanent address:
Institute for Theoretical Physics, Kanazawa 
University}
\vspace*{5mm}\\
{\em 
Max-Planck-Institut f\"ur Physik,
 Werner-Heisenberg-Institut\\
D-80805 Munich, Germany}\\
\end{center}
\vspace*{50mm}
\begin{abstract}
The leptonic sector in a  recently proposed 
minimal extension  of the
standard model, in which the permutation symmetry
$S_3$ is assumed to be an exact flavor symmetry at the weak scale,
is revisited.
We find that 
$S_3$ 
 with an additional $Z_{N}$ symmetry
 allows   CP violating  phases
in  the neutrino mixing.
 The leptonic  sector contains
six real parameters with two independent phases
to describe  charged lepton and neutrino masses and  the neutrino mixing.
The model predicts:
an inverted spectrum of neutrino mass, 
$\tan\theta_{23}=1+O(m_e^2/m_\mu^2)$ 
and $\sin\theta_{13}=m_e/\sqrt{2}m_\mu+O(m_em_\mu/m_{\tau}^2)
\simeq 0.0034$.
Neutrino mass as well as the effective Majorana mass $<m_{ee}>$
in the neutrinoless double-$\beta$ decay can be expressed in a closed form
as a function of 
 $\phi_\nu, \Delta m^2_{21},\Delta m^2_{23}$ and $\tan\theta_{12}$,
 where $\phi_\nu$ is one of the  independent phases.
The model also predicts
$<m_{ee}> \geq ( 0.036  - 0.066  )$ eV.

\end{abstract}
\vspace*{20mm}

\newpage
\pagestyle{plain}
\pagenumbering{arabic}
\setcounter{footnote}{0}

The Yukawa sector of the standard model (SM),
which is responsible for the generation of  the mass of leptons and
quarks, and their mixing,
has too many redundant parameters.
This not only weakens the predictivity of the SM, but also 
makes ambiguous how to go beyond the SM.
An exact flavor 
symmetry could reduce this redundancy,
thereby  giving useful 
hints about how to unify the flavor structure of  the SM.

Recently, a minimal $S_3$ invariant extension of the
SM was suggested in  \cite{kubo}, while assuming that 
the Higgs, quark and  lepton including the
right-handed neutrino fields belong to the three-dimensional
reducible representation of the permutation group $S_3$
 \footnote{A
partial list for  permutation symmetries is
\cite{pakvasa1}--\cite{ma4}.
See for instance
\cite{fritzsch} for a review. The basic idea of \cite{kubo}
is similar to that of \cite{pakvasa1,koide2,ma}.}.
This  smallest nonabelian 
symmetry based on  $S_3$ 
 is only spontaneously broken, because the electroweak gauge symmetry
$SU(2)_L\times U(1)_Y$ is spontaneously broken.
It was found in \cite{kubo} that
this flavor symmetry 
is consistent with experiments, and that
in the leptonic sector
an additional discrete symmetry $Z_2$ can be introduced.
It was argued there that due the additional discrete $Z_2$ symmetry
the neutrino mixing matrix $V_{\rm MNS}$ can not contain
any CP violating phase \footnote{See for instance \cite{grimus-r2}
for recent reviews on CP violation in the leptonic sector.}.
We now believe this
is incorrect, and   we would like to re-investigate
the leptonic sector of the model in this letter.

We will find that  it is possible to introduce 
two independent
 CP violating  phases \cite{bilenky}
in  the neutrino mixing
 even with
an additional $Z_{N}$ symmetry in the leptonic sector.
The permutation symmetry $S_3$
with  $Z_{N}$  allows
three real mass parameters for the
charged lepton mass matrix, and three real parameters 
and two phases
for the neutrino mass matrix.
The model predicts \footnote{Similar but different predictions 
are obtained from
different types of discrete symmetry  \cite{fukuyama}--\cite{grimus2}.
See also \cite{xing} and  \cite{ma4}.}:
an inverted spectrum of neutrino mass, 
$\tan\theta_{23}=1+O(m_e^2/m_\mu^2)$ 
and $\sin\theta_{13}=m_e/m_\mu\sqrt{2}+O(m_e m_\mu/m_{\tau}^2)$.
Neutrino mass as well as the effective Majorana mass $<m_{ee}>$
in the neutrinoless double-$\beta$ decay can be expressed in a closed form
as a function of 
 $\phi_\nu, \Delta m^2_{21},\Delta m^2_{23}$ and $\tan\theta_{12}$,
 where $\phi_\nu$ is one of the  independent phases.
We find that the minimum  of $m_{\nu_2}$ as well as $<m_{ee}>$
occurs at $\phi_\nu=0$, which is  approximately 
$\sqrt{\Delta m^2_{23}}/\sin 2\theta_{12}$.
 
Before we will come to our main purpose of the letter, let us briefly
summarize the basic ingredient of the $S_3$ invariant SM of \cite{kubo}.
The quark, lepton and Higgs fields are denoted by
$Q^T=(u_L,d_L)~,~ u_R~,~d_R~,~L^T=(\nu_L,e_L)~,~e_R~,~ 
\nu_R~,~H$. 
Each of them forms
a reducible representation ${\bf 1}+{\bf 2}$ of $S_3$.
The doublets carry capital indices $I,J$ which run from $1$ to $2$,
and the singlets are denoted by
$Q_3 , u_{3R},u_{3R},L_3~,~
e_{3R}~,~\nu_{3R}~,~H_3$.
The most general renormalizable Yukawa interactions are given by
\be
{\cal L}_Y &=& {\cal L}_{Y_D}+{\cal L}_{Y_U}
+{\cal L}_{Y_E}+{\cal L}_{Y_\nu},
\label{lag}
\ee
where
\be
{\cal L}_{Y_D} &=&
- \overline{ Q} ~\sum_{i=1}^3 {\bf Y}_{H_i}^{d} H_i~d_{R}+h.c. ,
\nn\\
{\cal L}_{Y_U} &=&
-\overline{ Q}(i \sigma_2)~\sum_{i=1}^3 {\bf Y}_{H_i}^{u} H_i ~u_{R}
 +h.c.,
\nn
\\
{\cal L}_{Y_E} &=& -\overline{ L}~\sum_{i=1}^3 {\bf Y}_{H_i}^{e} H_i~
 e_{R} +h.c.,
\nn\\
{\cal L}_{Y_\nu} &=& -\overline{ L} (i \sigma_2) 
~\sum_{i=1}^3{\bf Y}_{H_i}^{\nu} H_i~\nu_{R}+h.c.,\nn
\ee
and the Yukawa coupling matrices are given by \cite{kubo}
\be
{\bf Y}_{H_1}^{k} &=& \left( \begin{array}{ccc}
0 & Y_2^{k} & Y_5^{k}
\\ Y_2^{k} & 0 & 0
  \\ Y_4^{k} & 0&  0
\end{array}\right)~,~
{\bf Y}_{H_2}^{k}= \left( \begin{array}{ccc}
Y_2^{k} & 0 & 0
\\  0 & -Y_2^{k} &Y_5^{k}
  \\ 0 & Y_4^{k}& 0
\end{array}\right),\\
{\bf Y}_{H_3}^{k} &=& \left( \begin{array}{ccc}
Y_1^{k} & 0&0
  \\  0& Y_1^{k} & 0
  \\ 0 & 0 &  Y_3^{k}
\end{array}\right)~,~k=d,u,l,\nu.
\ee
Further,  the Majorana mass terms for the right-handed neutrinos is given by
\be
{\cal L}_{M} =
-M_1 \nu_{IR}^T C \nu_{IR} 
-M_3 \nu_{3R}^T C \nu_{3R},
\label{majorana}
\ee
where $C$ is the charge conjugation 
matrix\footnote{Supersymmetrization of the present model
has been proposed in \cite{kobayashi1}.}.

Pakvasa and Sugawara \cite{pakvasa1} analyzed
the Higgs potential.
The potential they analyzed has 
not only an abelian discrete symmetry (which we will use
for selection rules of the Yukawa couplings), but also
a permutation symmetry
$S_{2}$:
$H_{1} \leftrightarrow H_{2}$, which is not a subgroup of 
the flavor group $S_3$ of the model. 
We 
assume throughout this letter
 that the vacuum can
respect this accidental symmetry of the Higgs potential, and 
\be
<H_{1}> &=& <H_{2}>
\ee
is satisfied.  [$<H_{1}>=-<H_{2}>$ would yield the same physics.]
Then  the Yukawa interactions 
(\ref{lag})
yield the mass matrices of the
general form
\be
{\bf M} = \left( \begin{array}{ccc}
m_1+m_{2} & m_{2} & m_{5} 
\\  m_{2} & m_1-m_{2} &m_{5}
  \\ m_{4} & m_{4}&  m_3
\end{array}\right).
\label{general-m}
\ee
The  Majorana mass for $\nu_L$ can be obtained from 
the see-saw mechanism \cite{yanagida},
and the corresponding mass matrix is given by $
{\bf M_{\nu}} = {\bf M_{\nu_D}}\tilde{{\bf M}}^{-1} 
({\bf M_{\nu_D}})^T$,
where $\tilde{{\bf M}}=\mbox{diag}(M_1,M_1,M_3)$.
The mass matrices are diagonalized by the unitary matrices $U'$s
as
$$
U_{d(u,e)L}^{\dag}{\bf M}_{d(u,e)}U_{d(u,e)R}  ~,~
U_{\nu}^{T}{\bf M_\nu}U_{\nu}.$$
The diagonal masses  can be complex, and so the physical
masses are their absolute values, which we denote by $m_{\nu_1}, m_{\nu_2}, m_{\nu_3},
m_{e}, m_{\mu},m_{\tau},$ etc.

It would be certainly desirable to classify,
in a similar way  as in \cite{ramond,haba1}, all  possible  mass matrices 
that are consistent with an additional discrete
abelian symmetry and with experimental data.
We, however,  leave this program to feature 
work.
Here we simply adopt the result of \cite{kubo}
that 
\be
Y_1^e &=& Y_3^e=Y_1^\nu=Y_5^\nu=0,
\label{y0}
\ee
and consequently
\be
m_1^e &=& m_3^e=m_1^\nu=m_5^\nu=0
\ee
follows from a $Z_2$ symmetry. We emphasize that
there are a number of
different charge assignments of $Z_N$ that can yield
(\ref{y0}) \footnote{
We do not consider $U(1)$ to avoid the appearance
of a (nearly) massless particle.}: Provided that the charge 
of $H_3$, $Q(H_3)$,  is different from 
$Q(H_{1,2})$, only the conditions
\be
Q(L_3) &=&Q(L_{1,2})=
Q(e_{3R})+Q(H_{1,2})=Q(e_{1,2R})+Q(H_{1,2})\nn\\
&=&Q(\nu_{1,2R})-Q(H_{1,2})=Q(\nu_{3R})-Q(H_{3})
\ee
modulo $N$  should be satisfied 
to forbid $Y_1^e, Y_3^e, Y_1^\nu$ and $Y_5^\nu$.
Unfortunately, none of the abelian discrete symmetries above 
is a symmetry in the quark sector. 
Note that if $Z_N$ is chiral, it is broken by QCD anyway
($S_3$ is not broken by  QCD,
because it is not a chiral symmetry.)
The symmetry violating effect of the quark sector
appears first at the two-loop level in the leptonic sector, so that 
the violation of $Z_N$  in the leptonic
sector may be assumed to be negligibly small. Therefore, we throughout 
neglect that violating effect \footnote{That is, we assume that the relation (\ref{y0}) 
is satisfied at the weak scale. If one assumes that it is satisfied at some higher scale,
one should take into account the renormalization group running of the parameters
\cite{haba2}. See also \cite{lindner} for further references.
We however expect that the corrections will be small 
in the present model, in contrast to models,
in which a large neutrino mixing is not related to a symmetry of the theory.}.

To proceed with our discussion, we calculate the unitary matrix $U_{eL}$  from
\be
U_{eL}^{\dag} {\bf M}_e{\bf M}_e^{\dag} U_{eL} =
\mbox{diag}(m_e^2, m_\mu^2, m_\tau^2),
\ee
where
\be
{\bf M}_e{\bf M}_e^{\dag}
&=&\left( \begin{array}{ccc}
2 (m^e_{2})^2+(m_{5}^e)^2 &  
(m_{5}^e)^2&  2 m^e_{2} m_{4}^e
\\  (m_{5}^e)^2 &2 (m^e_{2})^2
+(m_{5}^e)^2& 0
  \\  2 m^e_{2} m_{4}^e & 0 &
 2 (m_{4}^e)^2
\end{array}\right),
\label{ml2}
\ee
and all the mass parameters appearing in (\ref{ml2}) are
real. We find that $U_{eL}$ can be approximately written as \cite{kobayashi1} 
\be
U_{eL} & \simeq &\left(
\begin{array}{ccc}
-\frac{y}{2}\left( 1 + \frac{1}{x^2}\right) &
-\frac{1}{\sqrt{2}} \left(1-\frac{y^2}{4}-\frac{y^2}{2x^2} \right) &
\frac{1}{\sqrt{2}} \\
\frac{y}{2}\left( 1 - \frac{1}{x^2}\right) &
\frac{1}{\sqrt{2}} \left(1-\frac{y^2}{4}+\frac{y^2}{2x^2} \right) &
\frac{1}{\sqrt{2}} \\
1-\frac{y^2}{4} & -\frac{y}{\sqrt{2}} &  \frac{y}{\sqrt{2}x^2}
\end{array}\right),
\label{UeL}
\ee
where  $x =m_5^e/m_2^e\simeq m_{\tau}/m_{\mu}$ and
$y  =m_4^e/m_2^e\simeq \sqrt{2}m_e/m_{\mu}$.

The Majorana masses of the right-handed neutrinos,
  $M_{1}$ and $M_{3}$ in (\ref{majorana})
which may be complex,  can be absorbed by a redefinition of
$m^{\nu}_{2}, m^{\nu}_{4}$ 
and $m^{\nu}_{3}$, and we may therefore
assume that   $M_{1}$ and $M_{3}$ are real.
After  rescaling of $m^{\nu}_{2}, m^{\nu}_{4}$ 
and $m^{\nu}_{3}$ as
\be
(m_2^\nu) &\to & \rho_2^\nu =(m_2^\nu) /M_1^{1/2}~,~
(m_4^\nu) \to \rho_4^\nu = (m_4^\nu) /M_1^{1/2}~,~
(m_3^\nu) \to  \rho_3^\nu =(m_3^\nu)/ M_3^{1/2},
\label{rescale}
\ee
we obtain
\be
{\bf M}_{\nu} & = &{\bf M_{\nu_D}}\tilde{{\bf M}}^{-1} 
({\bf M_{\nu_D}})^T=
\left( \begin{array}{ccc}
2 (\rho^{\nu}_{2})^2 & 0 & 
2 \rho^{\nu}_2 \rho^{\nu}_{4}
\\ 0 & 2 (\rho^{\nu}_{2})^2 & 0
  \\ 2 \rho^{\nu}_2 \rho^{\nu}_{4} & 0  &  
2 (\rho^{\nu}_{4})^2 +
(\rho^{\nu}_3)^2
\end{array}\right).
\label{m-nu}
\ee
All the phases in (\ref{m-nu}), 
except for one,  can  be  absorbed.
Without loss of generality, we may assume  that $\rho_3^\nu $ is complex.
We find that ${\bf M}_{\nu}$ can be diagonalized as
\be
U^T_\nu {\bf M}_{\nu} U_\nu &=& \left( \begin{array}{ccc}
m_{\nu_1}e^{i\phi_1-i\phi_\nu} & 0 & 0\\
0 & m_{\nu_2}e^{i\phi_2+i\phi_\nu} &0 \\
0 & 0 & m_{\nu_3}
\end{array}\right),
\label{unu}
\ee
where
\be
U_{\nu}&= &\left( \begin{array}{ccc}
-s_{12} & c_{12}e^{i \phi_\nu}
&  0
\\ 0 & 0 &1
\\    c_{12}e^{-i \phi_\nu}  & s_{12}& 0
 \end{array}\right),\\
\label{unumax3}
m_{\nu_3} \sin \phi_\nu &=& m_{\nu_2} \sin \phi_2
=m_{\nu_1} \sin \phi_1,
\label{sinp}
\ee
and $c_{12}=\cos\theta_{12}$ and $s_{12}=\sin\theta_{12}$.
The mixing angle is given by
\be
\tan^2\theta_{12} &=&
\frac{(m_{\nu_2}^2-m_{\nu_3}^2 \sin^2\phi_\nu)^{1/2}
-m_{\nu_3}|\cos\phi_\nu|}{(m_{\nu_1}^2
-m_{\nu_3}^2 \sin^2\phi_\nu)^{1/2}
+m_{\nu_3}|\cos\phi_\nu|},
\ee
from which we find
\be
\frac{m_{\nu_2}^2}{\Delta m_{23}^2} &=&
\frac{(1+2 t_{12}^2+t_{12}^4-r t_{12}^4)^2}
{4  t_{12}^2 (1+t_{12}^2)(1+t_{12}^2-r t_{12}^2)\cos^2 \phi_\nu}
-\tan^2 \phi_\nu\\
&\simeq &
\frac{1}{\sin^2 2\theta_{12}\cos^2 \phi_\nu}
-\tan^2 \phi_\nu ~~\mbox{for}~~|r| << 1,
\label{mnu2}
\ee
where $t_{12}=\tan\theta_{12}, r=\Delta m_{21}^2/\Delta m_{23}^2$.
As in \cite{kubo}, we find that  only an inverted mass spectrum
\be
m_{\nu_3} & < & m_{\nu_1}, m_{\nu_2}
\label{spectrum}
\ee
is consistent with  the experimental constraint $ |\Delta m_{21}^2|
< |\Delta m_{23}^2|$  in the present model.
To see this, we first derive
\be
m_{\nu_1}\cos \phi_1-m_{\nu_3}\cos\phi_\nu
&=&-2\rho_2^\nu \rho_4^\nu ~A_1\\
m_{\nu_2}\cos\phi_2-m_{\nu_3}\cos\phi_\nu
&=& 2\rho_2^\nu \rho_4^\nu~A_2,
\ee
where
\be
A_1 &=& \sin 2\theta_{12}
+\cos^2 \theta_{12}/\tan 2 \theta_{12}~,~
 A_2 = \sin 2\theta_{12}-\sin^2 \theta_{12}/\tan 2 \theta_{12}.
\ee
Then we use the fact that if $A_1$ is positive (negative),
then $A_2$ is always positive (negative).
Suppose that $2\rho_2^\nu \rho_4^\nu~A_2$ is positive, which
 implies that $m_{\nu_2}\cos\phi_2 > m_{\nu_3}\cos\phi_\nu$ 
 and $m_{\nu_1}\cos\phi_1 < m_{\nu_3}\cos\phi_\nu$.
In this case, eq. (\ref{sinp}) can be satisfied, only if 
$m_{\nu_2} > m_{\nu_3}$ or $m_{\nu_1} > m_{\nu_3}$.
Similarly, if $-2\rho_2^\nu \rho_4^\nu~A_1$ is positive,
then $m_{\nu_2} > m_{\nu_3}$ or $m_{\nu_1} > m_{\nu_3}$
 has to be satisfied.
Therefore, $m_{\nu_3}$ cannot be the largest among
$m_{\nu_i}$'s
\footnote{
Of course, $m_{\nu_1} > m_{\nu_3}>m_{\nu_2}$ or
$m_{\nu_2} > m_{\nu_3}>m_{\nu_1}$ 
 is mathematically allowed, but is excluded by experiments.}.

In fig.~1 we  plot $m_{\nu_2}$  versus
$\sin\theta_{12}$ for  $ \Delta m_{21}^2=6.9 \times10^{-5}$ eV$^2$,
$\Delta m_{23}^2=2.5\times 10^{-3}$ eV$^2$ 
(best-fit values reported in \cite{maltoni,fogli,pakvasa2}) and
 $\sin \phi_\nu=0$ (solid), $0.6$ (dotted) and $0.96$ (dashed).
The $\sin\phi_\nu$ dependence of 
$m_{\nu_2}$ is shown in fig.~2 for 
$\tan\theta_{12}=0.68$ and the same values of 
$ \Delta m_{21}^2$ and 
$\Delta m_{23}^2$ as in fig.~1.
As we see from (\ref{mnu2}) and also from fig.~2, $m_{\nu_2}$
assumes at $\sin\phi_\nu=0$ its minimal value
\be
m_{\nu_2, {\rm min}} &\simeq  \sqrt{\Delta m_{23}^2}/\sin2\theta_{12}
=   ( 0.036  - 0.066  ) ~~\mbox{eV}
\label{nu2min},
\ee
where we have used 
$\Delta m_{23}^2=( 1.3  -  3.0 )\times 10^{-3}$ eV$^2$
and $\sin 2\theta_{12}=0.83-1.0$ \cite{maltoni}--\cite{nishikawa}.

\begin{figure}[tb]
\includegraphics*[width=1\linewidth]
{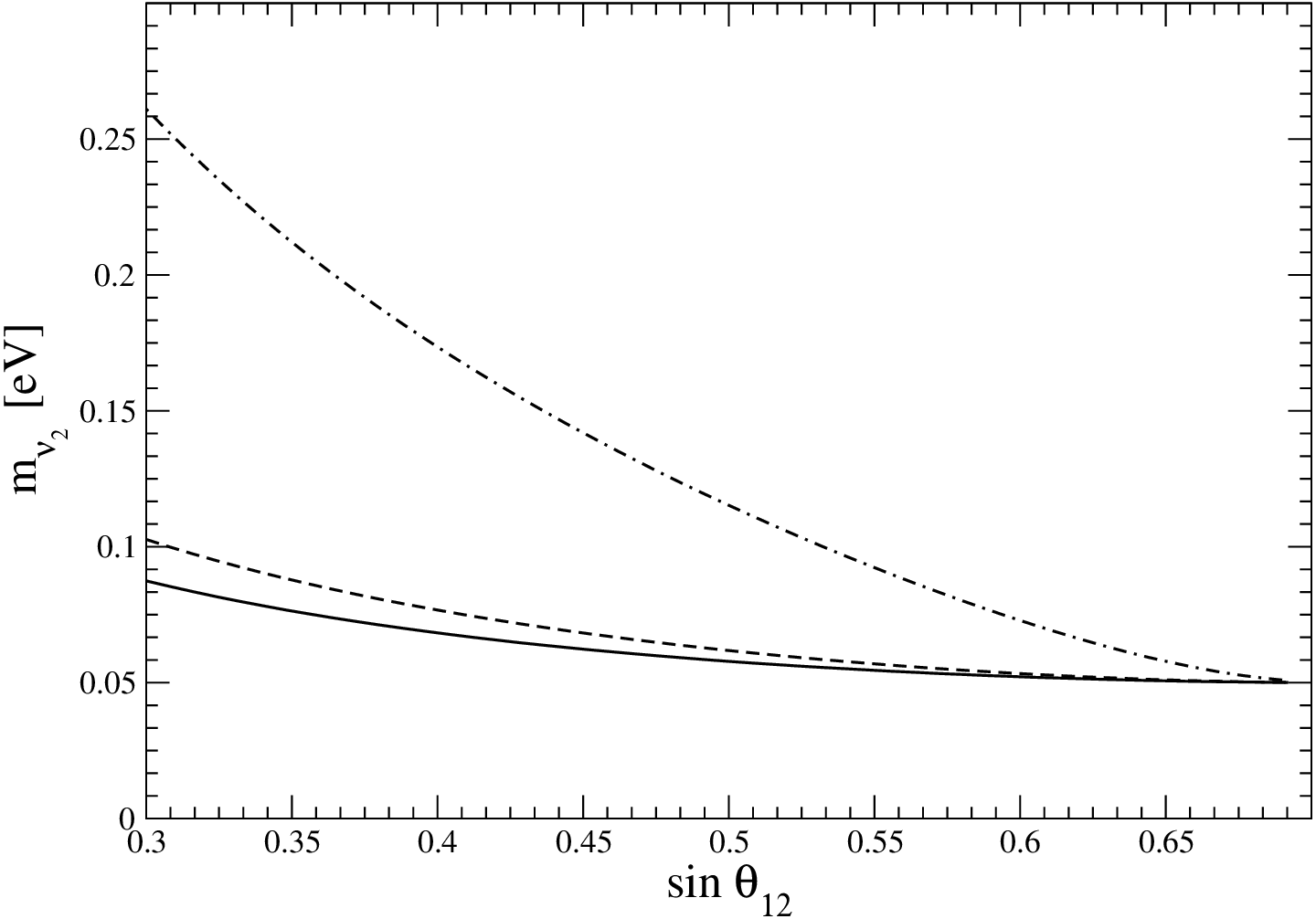}
\caption{$m_{\nu_2}$  versus $\sin\theta_{12}$ 
 for  $ \Delta m_{21}^2=6.9 \times10^{-5}$ eV$^2$,
$\Delta m_{23}^2=2.5\times 10^{-3}$ eV$^2$ and
 $\sin \phi_\nu=0$ (solid), $0.6$ (dotted) and $0.96$ (dot-dashed).}
\label{fig1}
\end{figure}
 
 \begin{figure}[tb]
\centerline{\includegraphics*[width=1\linewidth]
 {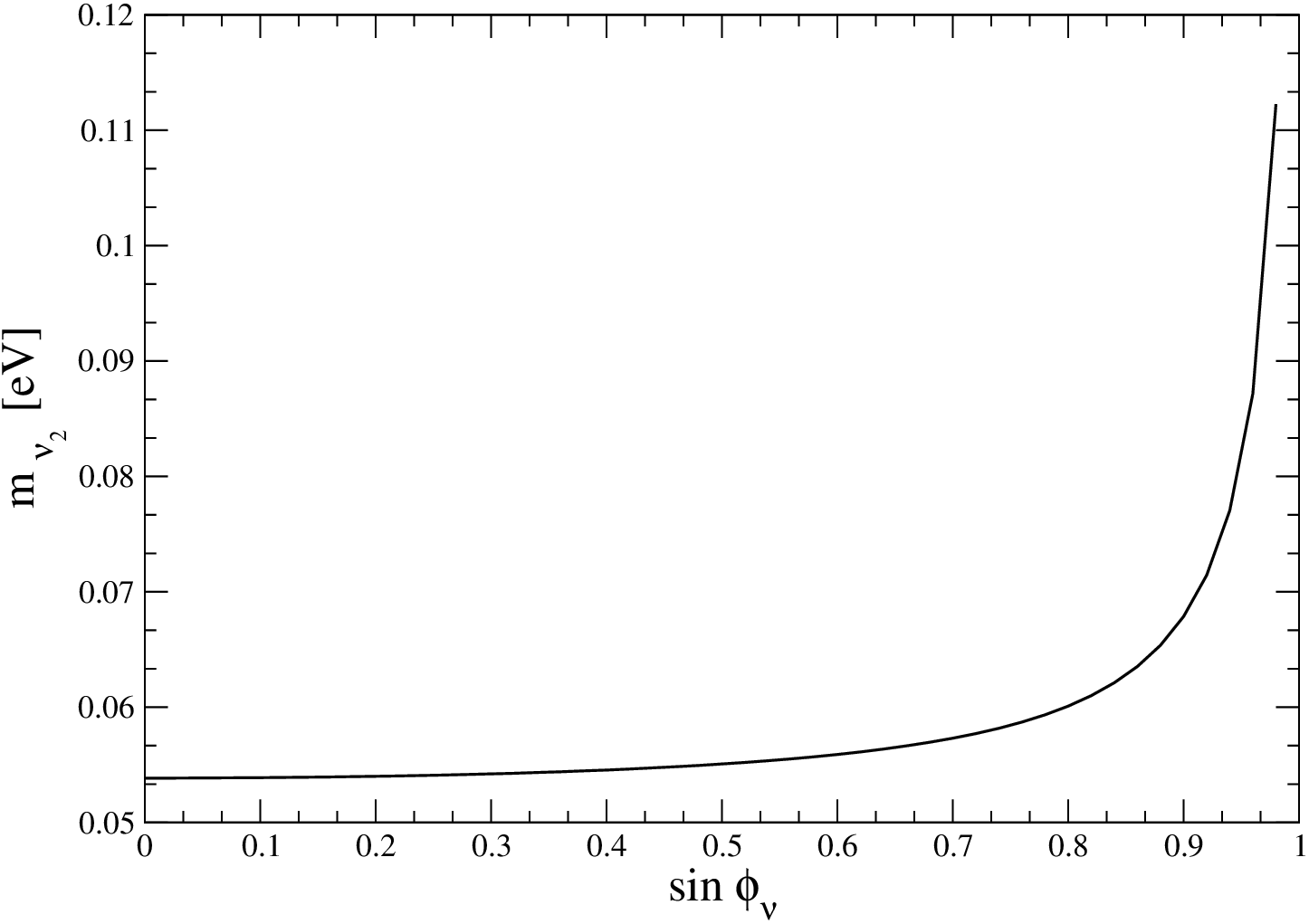}}
\caption{$m_{\nu_2}$ as a function of  $\sin\phi_\nu$ 
 for 
$\tan\theta_{12}=0.68, \Delta m_{21}^2=6.9 \times10^{-5}$ eV$^2$ and 
$\Delta m_{23}^2=2.5\times 10^{-3}$ eV$^2$.}
\label{fig2}
\end{figure}

Now the product $U_{eL}^{\dag} P  U_\nu$ 
with $P=\mbox{diag.}(1,1,\exp i \mbox{arg} (Y_4^\nu))$
defines a neutrino mixing matrix, which we bring by an
appropriate  phase transformation  to the popular form
\be
V_{\rm MNS} &=&
 \left( \begin{array}{ccc}
 c_{12} c_{13} & s_{12} c_{13} &   s_{13} e^{-i\delta}\\
  -s_{12} c_{23}  -c_{12} s_{23} s_{13}e^{i\delta} & 
   c_{12} c_{23}  -s_{12} s_{23} s_{13}e^{i \delta}  &   s_{23}c_{13} \\
 s_{12} s_{23}  -c_{12} c_{23} s_{13}e^{i \delta} & 
   -c_{12} s_{23}  -s_{12} c_{23} s_{13}e^{i \delta}  &   c_{23}c_{13}
\end{array}\right) 
\left( \begin{array}{ccc}
1 & 0 & 0\\
0 & e^{i \alpha} &0 \\
0 & 0 & e^{i \beta}
\end{array}\right).
\ee
We find:
\be
s_{13} & =& \frac{1}{\sqrt{2}}\frac{m_e}{m_\mu}
+O(m_e m_\mu/m_{\tau}^2) \simeq 0.0034,~
t_{23} = \frac{s_{23}}{c_{23}}=
1-\frac{1}{2}(\frac{m_e}{m_\mu})^2+O(m_e^2 /m_{\tau}^2),
\label{s123}\\
\delta &=&\mbox{arg} (Y_4^\nu)-\phi_\nu ,\\
\sin 2 \alpha &=&\sin(\phi_1-\phi_2)\nn\\
& =&
\pm \frac{ m_{\nu_3}\sin\phi_\nu}{m_{\nu_1}m_{\nu_2}}
\left( \sqrt{m_{\nu_2}^2-m_{\nu_3}^2 \sin^2 \phi_\nu}+
\sqrt{m_{\nu_1}^2-m_{\nu_3}^2 \sin^2 \phi_\nu} \right)
\label{alpha}\\
&\simeq &  \pm 2 \sin\phi_\nu (m_{\nu_3}/m_{\nu_2})
\sqrt{1-(m_{\nu_3}/m_{\nu_2})^2 \sin^2\phi_\nu},\\
\sin 2 \beta &=&\sin(\phi_1-\phi_\nu)\nn\\
& = &
\pm \frac{\sin\phi_\nu}{m_{\nu_1}}
\left(m_{\nu_3}
 \sqrt{1-\sin^2 \phi_\nu}+
 \sqrt{m_{\nu_1}^2-m_{\nu_3}^2 \sin^2 \phi_\nu}
\right ),
\label{sinb}
\ee
for $\phi_1+\phi_2 \sim \pm \pi$,
where $\phi_1,\phi_2$ and $\phi_\nu$ are defined in (\ref{unu}).
 Since $\sin^2 2 \theta_{13}
\simeq 4.6\times 10^{-5}$, future oscillation
experiments such as J-Park experiment  \cite{nishikawa} can easily exclude the model.
In fig.~3 we  plot $\sin 2 \alpha$ (solid) and
$\sin 2 \beta$ (dotted) as a function of $\sin \phi_\nu$.
As we can see, $\sin 2 \alpha$ reaches its maximal value
$1$ at $\sin \phi_\nu \simeq 0.94$.
Similarly, the maximal value of $\sin 2 \beta$,
which is about $1$, 
occurs  at $\sin \phi_\nu \simeq 0.85$.
\begin{figure}
\centerline{\includegraphics*[width=1\linewidth]
{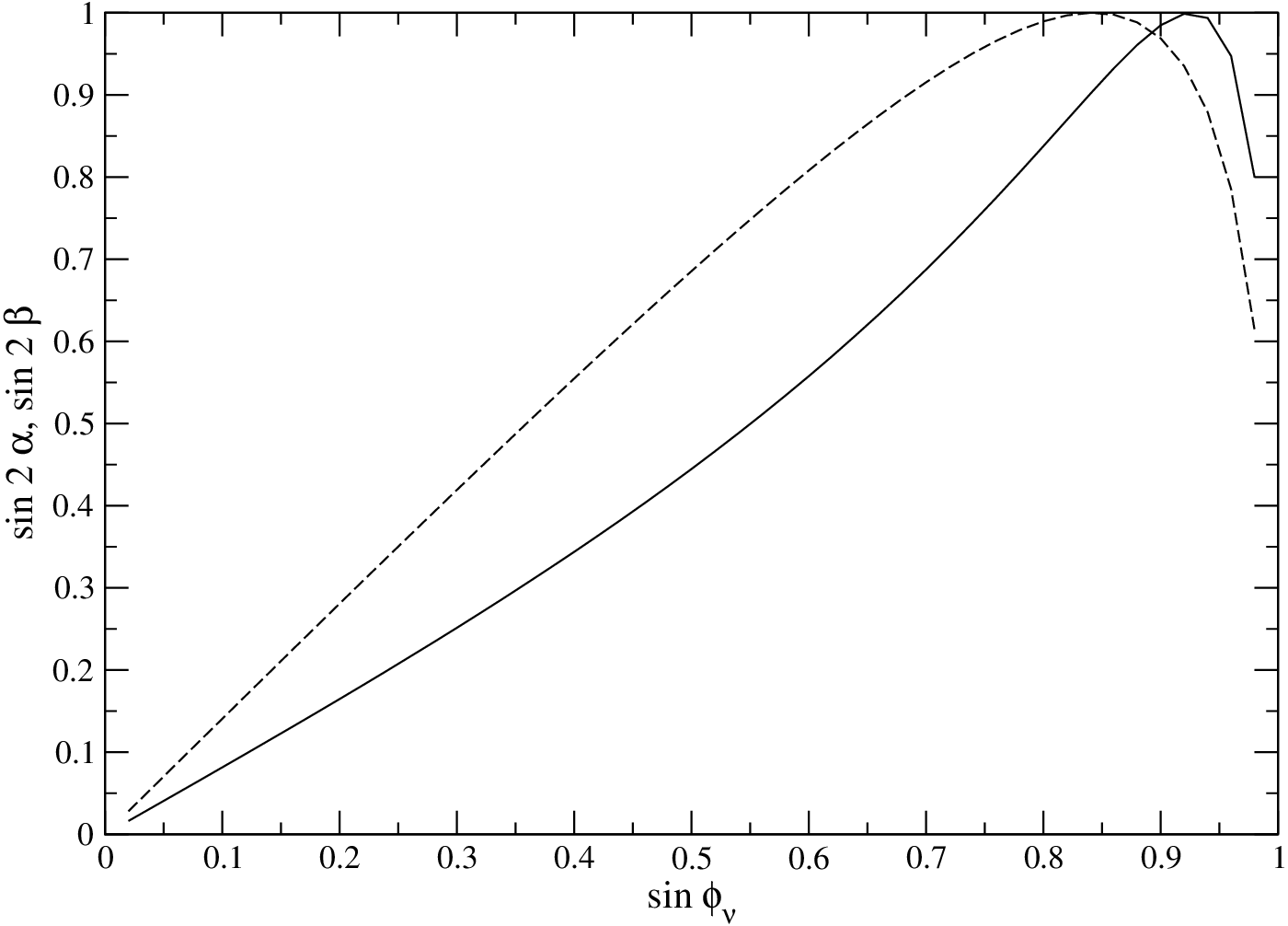}}
\caption{$\sin 2 \alpha$ (solid) and
$\sin 2 \beta$ (dotted) versus $\sin \phi_\nu$
for $\tan\theta_{12}=0.68, \Delta m_{21}^2=6.9 \times10^{-5}$ eV$^2$ and
$\Delta m_{23}^2=2.3 \times 10^{-3}$ eV$^2$
in the case of $\phi_1+\phi_2 \sim \pi$.}
\label{fig3}
\end{figure}
\begin{figure}
\centerline{\includegraphics*[width=1\linewidth]
{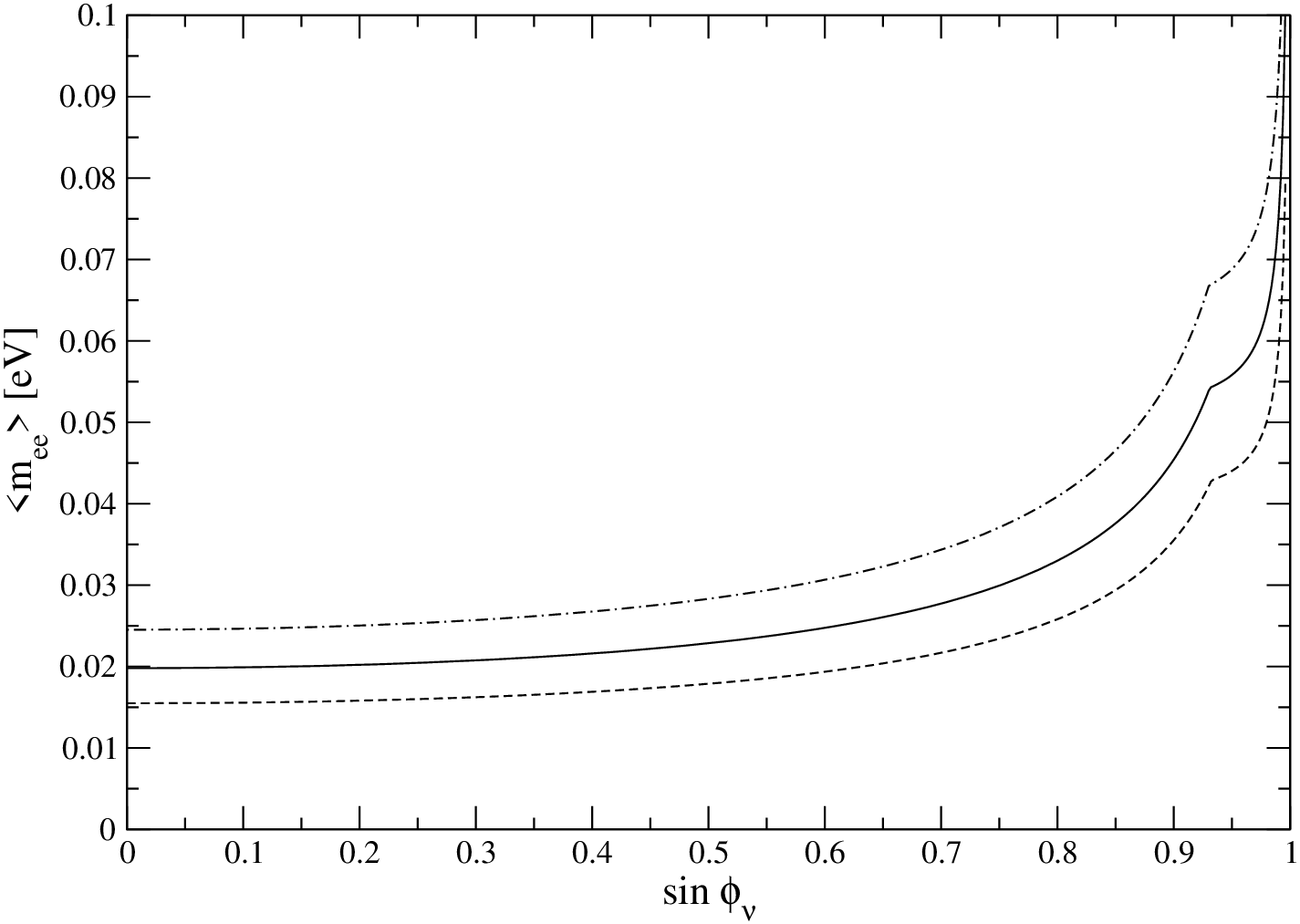}}
\caption{The effective Majorana mass $<m_{ee}>$ as a function of
$\sin \phi_\nu$ with 
$\sin^2\theta_{12}=0.3$ and 
 $\Delta m_{21}^2=6.9 \times10^{-5}$ eV$^2$.
 The dashed, solid and dot-dashed lines stand for 
$\Delta m_{23}^2=1.4, 2.3$ and $ 3.0 \times 10^{-3}$ eV$^2$,
respectively. The $\Delta m_{21}^2$ dependence is very small.}
\label{fig4}
\end{figure}
We then consider the effective Majorana mass
\be
<m_{ee}> &=& |~\sum_{i=1}^3m_{\nu_i} V_{ei}^2|
\simeq |m_{\nu_1}c_{12}^2+
m_{\nu_2}s_{12}^2 \exp i2\alpha~ |,\\
&\simeq &\frac{\sqrt{\Delta m_{23}^2}}{\cos\phi_\nu}~
~[\sin^{-2}2\theta_{12}-\sin^2\phi_\nu]^{1/2}~
[1+\frac{\sin^2 2\theta_{12}}{2}(\cos 2\alpha-1)]^{1/2},
\label{mee}
\ee
which can be measured in neutrinoless double $\beta$ decay
experiments. ($\alpha$ is given in (\ref{alpha}).)
In fig.~4 we  plot $<m_{ee}>$  as a function of $\sin \phi_\nu$.
As we can see from fig.~4, the effective Majorana mass stays
at about its minimal value $<m_{ee}>_{\rm min}$ for a wide range of $\sin\phi_\nu$.
Since $<m_{ee}>_{\rm min}$ is 
approximately equal to $m_{\nu_2,{\rm min}}$
(which is given in (\ref{nu2min})), it is consistent with recent experiments \cite{spergel,klapdor1}
and is within an accessible range of future experiments \cite{klapdor2}.
An experimental verification of (\ref{mnu2}), (\ref{spectrum}) and 
(\ref{s123})--(\ref{mee}) would strongly indicate the existence of
the smallest nonabelian symmetry based on the permutation group
$S_3$ along with an abelian discrete symmetry $Z_N$ at the electroweak
scale, where  $Z_N$ is only an approximate symmetry of the whole theory,
but the effect of its violation
is of two-loop order in the leptonic sector.

$S_3$ is obviously a possible answer to the question why there exist three
generations of  leptons and quarks.
$S_3$, of course, can not explain the hierarchy of the 
fermion mass spectrum, but $S_3$ with $Z_N$
in the leptonic sector
can relate the mass spectrum and  mixing in this sector, making testable predictions,
which have been re-investigated in the present letter.
Therefore, $S_3$ solves partially the flavor problem of the SM.
Since there are three $SU(2)_L$ doublet Higgs fields
in the  model, there exit FCNC processes at the tree level.
In \cite{kubo} the magnitude of various  tree level FCNC amplitudes have
been estimated, and it has been found that they 
are sufficiently suppressed. 
The suppression follows from the smallness of
the corresponding Yukawa couplings, where $S_3$ plays an 
important role for that smallness.
However, we find that $\Delta m_K$, the difference of the mass of $K_L$ and $K_S$,
exceeds the experimental value, unless the mixing of the Higgs fields
is fine tuned. This problem is currently under investigation, and
we will report the result elsewhere.

It is straightforward to keep
the discrete flavor symmetries,
$S_3$ in the hadronic sector and $S_3 \times Z_N$
in the leptonic sector, in a
supersymmetric extension of
the standard model \cite{kobayashi1}.
The supersymmetric flavor problem has been  investigated there, and 
it has been explicitly  found that thanks to
the flavor symmetries the dangerous
FCNC and CP violating  processes,
that originate from  soft supersymmetry breaking terms,  are
sufficiently suppressed, in a similar manner as it was found in \cite{hamaguchi}.

\vspace{0.5cm}
\noindent
{\large \bf Acknowledgments}\\
This work is supported by the Grants-in-Aid for Scientific Research 
from the Japan Society for the Promotion of Science
(No. 13135210).

\newcommand{\bi}{\bibitem}

\end{document}